\documentclass[10pt,a4paper]{article}
\usepackage{amssymb, amsmath, amsthm }
\usepackage{graphicx,subfig}
\usepackage{setspace, bbm}

\newcommand{\argmax}[1]{\underset{#1}{\operatorname{argmax}}}
\newcommand{\ba}{\begin{align*}}
\newcommand{\eaa}{\end{align*}}

\newcommand{\nl}{\notag\\}

\newcommand{\frn}{\frac 1 n}

\newcommand{\half}{\frac 1 2}

\newcommand{\calX}{{\cal X}}

\newcommand{\calY}{{\cal Y}}
\newcommand{\calZ}{{\cal Z}}
\newcommand{\calU}{{\cal U}}
\newcommand{\calV}{{\cal V}}

\newcommand{\calA}{{\cal A}}

\newcommand{\calW}{{\cal W}}

\newcommand{\calQ}{{\cal Q}}

\newcommand{\calS}{{\cal S}}

\newcommand{\calR}{{\cal R}}
\newcommand {\bx} {\mbox{\boldmath $x$}}

\newcommand {\by} {\mbox{\boldmath $y$}}

\newcommand {\bE} {\mbox{\boldmath $E$}}

\newcommand {\hX}{\hat{X}}

\newcommand {\hx}{\hat{x}}

\newcommand {\hY}{\hat{Y}}

\newcommand{\eqd}{\stackrel{\triangle}{=}}
\newcommand {\exe} {\stackrel{\cdot} {=}}

\newcommand{\sbr}[1] {\left[#1\right]}
\newcommand{\cbr}[1] {\left\{#1\right\}}
\newcommand{\rbr}[1] {\left(#1\right)}
\newtheorem{theorem}{Theorem}

\newtheorem{corrolary}{Corrolary}

\onehalfspace
\begin{document}
\title{Zero-Delay and Causal Single-User and Multi-User Lossy Source Coding with Decoder Side Information\footnote{This research was supported by the Israeli Science Foundation (ISF), grant no. 208/08.}}
\author{Yonatan Kaspi and Neri Merhav\\
Department of Electrical Engineering \\
Technion - Israel Institute of Technology \\
Technion City, Haifa 32000, Israel\\
Email: \{kaspi@tx, merhav@ee\}.technion.ac.il}


\maketitle

\begin{abstract}
We consider zero-delay single-user and multi-user source coding with average distortion constraint and decoder side information. The zero-delay constraint translates into causal (sequential) encoder and decoder pairs as well as the use of instantaneous codes. For the single-user setting, we show that optimal performance is attained by time sharing at most two scalar encoder-decoder pairs, that use zero-error side information codes. Side information lookahead is shown to useless in this setting. We show that the restriction to causal encoding functions is the one that causes the performance degradation, compared to unrestricted systems, and not the sequential decoders or instantaneous codes.
Furthermore, we show that even without delay constraints, if either the encoder or decoder are restricted a-priori to be scalar, the performance loss cannot be compensated by the other component, which can be scalar as well without further loss. 
Finally, we show that the multi-terminal source coding problem can be solved in the zero-delay regime and the rate-distortion region is given.
\end{abstract}

\section{Introduction}
The classical source coding theorems and their converse counterparts provide fundamental limits which are usually asymptotic in the sense that they can be achieved by systems that introduce delay (imposed by operating on blocks) and/or require complexity that grows exponentially. In many practical scenarios, however, delay is not tolerable. Such systems include many real-time source coding applications such as streaming live multimedia. In such applications, not only should the data be encoded and decoded in real-time, but also there is no way to improve the reconstruction of previously decoded data as new data arrive. While zero-delay is clearly motivated by fast real-time systems, it can also be motivated by extremely slow systems that observe data once in a long while and must act on this data before new data will arrive. This work focuses on the fundamental limits of such zero-delay systems with the addition of side information (SI) which is available to the decoder.         

Zero-delay operation means that each time a new source symbol is observed, a message must be sent to the decoder. The decoder must decode the message and reconstruct the source symbol before the next message will arrive.  This, in turn, translates into three constraints: Primarily, the encoder functions must be causal functions of the source symbols. Secondly, the code with which the encoders sends the messages to the decoder must be instantaneous, meaning the decoder can detect the end of each codeword before the whole codeword of the next symbol will arrive. Alternatively, the decoder must be able to parse the bit-stream which is composed of the received codewords in a causal manner. Finally, the decoder must be a causal function of the encoder messages and its SI. 

In this work, we consider both single- and multi-user scenarios. We start by describing the single-user setting and related work and then continue to the multiterminal scenarios. In the single-user setting, we consider the following source coding problem: Symbols produced by a discrete memoryless source (DMS) are to be encoded, transmitted noiselessly and reproduced by a decoder without delay. The decoder has access to SI correlated to the source. The average distortion between the source and the reproduced symbols is constrained to be smaller than some predefined constant. 

When no distortion is allowed, this problem falls within the scope of zero-error source coding with SI, which was initially introduced by Witsenhausen in \cite{Witsenhausen1976}.  Witsenhausen considered fixed-length coding and characterized the side-information structure as a confusability graph defined on the source alphabet. With this characterization, fixed-length SI codes were equivalent to colorings of the associated graph. Alon and Orlitsky \cite{AlonOrlitski96} considered variable-rate codes for the zero-error problem. Two classes of codes were considered and lower and upper bounds were derived for both the scalar and infinite block length regimes. The work of Alon and Orlitsky was further extended by Koulgi \textit{et. al} \cite{KoulgiTuncel03}, who showed that the asymptotic zero-error rate of transmission is the complementary graph entropy of an associated graph. It was also shown in \cite{KoulgiTuncel03} that the design of optimal code is $NP$-hard and a sub-optimal, polynomial time algorithm was proposed. The combination of zero-error codes and maximum per-letter distortion was considered in \cite{TuncelDCC02}. As we will see in the sequel, the zero-error source coding with SI is relevant to our setting as well, since the encoder can rely on the SI when sending its messages. 

When the source alphabet is finite and distortion is allowed, scalar quantizer design boils down to finding the best partition of the source alphabet into disjoint subsets. The number of such subsets will be governed by the constraints imposed on the system (distortion, rate, encoder's output entropy etc.). In \cite{MuresanEffros08}, Muresan and Effros proposed an algorithm for finding good partitions in various settings which include the variable-rate scalar Wyner-Ziv \cite{WynerZiv76} setting. However, the optimality of the partitions relied on the convexity of the subsets. Namely, the subsets in each partition must be intervals in the source alphabet. It was noted by the authors that this requirement is too strong in the scalar Wyner-Ziv setting and there are many cases where the optimal partition contains non-convex subsets. We demonstrate such a scenario in the examples section of this work. Bounds on the performance of scalar, fixed-rate source codes with decoder SI were recently given in \cite{ReaniMerhav2012}.

Real--time codes form a subclass of the class of causal codes, as defined by Neuhoff and Gilbert \cite{NeuhoffGilbert1982}. In \cite{NeuhoffGilbert1982}, entropy coding is used on the whole sequence of reproduction symbols, introducing arbitrarily long delays. In the real time case, entropy coding has to be instantaneous, symbol--by--symbol (possibly taking into account past transmitted symbols). It was shown in \cite{NeuhoffGilbert1982} that for a DMS, the optimal causal encoder consists of time-sharing between no more than two scalar encoders. In \cite{TsachyNeri2005}, Weissman and Merhav extended \cite{NeuhoffGilbert1982} to include SI at the decoder, encoder or both. The discussion in \cite{TsachyNeri2005} was restricted, however, only to settings where the encoder and decoder could agree on the reconstruction symbol (i.e., the SI was used for compression, but not in the reproduction at the decoder). Non-causal coding of a source when the decoder has causal access to SI (with possibly a finite look-ahead) was considered by Weissman and El-Gamal \cite{TsachyElGamal06}. Gaarder and Slepian \cite{GaarderSlepian1982} gave structure theorems for fixed-rate encoders and decoders that are time-invariant finite-state devices. 

The results of \cite{NeuhoffGilbert1982} for causal coding can be adapted to zero-delay coding by replacing the arbitrary long delay entropy coding with zero-delay Huffman coding, thus showing that time-sharing at most two scalar quantizers, followed by Huffman coding, is optimal. When the SI is available to both the encoder and decoder, the results of \cite{TsachyNeri2005} can be adapted to zero-delay in a similar manner, where at most two scalar quantizers followed by Huffman coding are used for every possible SI symbol. The setting where the decoder can use the SI both to decode the compressed message and to reproduce the source was left open in  \cite{TsachyNeri2005}. As will be seen in the sequel, our results on zero-delay can be easily adapted back to the causal setting and answer some of the questions left open by  \cite{TsachyNeri2005}.

This paper has several contributions. In the single-user setting, the first contribution is the extension of \cite{NeuhoffGilbert1982} to zero-delay with decoder SI and average distortion constraint, where unlike \cite{TsachyNeri2005}, we do not restrict the usage of the SI. We show that results in the spirit of  \cite{NeuhoffGilbert1982} continue to hold here in the sense that it is optimal to time-share at most two scalar encoders and decoders. However, unlike the encoders in \cite{NeuhoffGilbert1982} that use Huffman codes in the zero-delay setting, here, the encoders transmit their messages using zero-error SI instantaneous codes, as defined in \cite{AlonOrlitski96} (and will be properly defined in the sequel). Secondly, we show that there is no performance gain if the decoder has non-causal access to the SI (lookahead) and in fact, only the current SI symbol is useful. This is in contrast to the arbitrary delay and causal SI setting of \cite{TsachyElGamal06}, where SI lookahead was shown to improve the performance. These results place the optimal performance of zero-delay systems far below the classical source coding results where arbitrary delay is allowed. We further ask which of the zero-delay constraints (causal encoder, instantaneous code and causal decoder) are causing this degradation in performance. It is shown that if we remove the constraint on the encoder and allow it to observe the whole sequence in advance but force instantaneous codes (restricting the number of bits in each transmission) and a causal decoder, at least in some cases, the classical rate-distortion performance can be obtained. This suggests that the ``blame'' for the relatively poor performance of the zero-delay systems falls, at least in these cases, on the restriction to causal encoding functions. The scheme we use to show the last point is surprisingly simple, but to the best of our knowledge it is novel nonetheless. Finally, we show that in the zero-delay setting, if we a-priori restrict attention to scalar decoders (that use only the current encoder message and SI symbol), scalar encoders will do as well as encoders that observe the whole source sequence in advance. Similarly, if we restrict the encoders to be scalar, scalar decoders will do as well as decoders that introduce delay and generally use all the encoder messages to reproduce the symbols. This means that the simplicity of one of the components (encoder / decoder) cannot be compensated by the complexity of the other component.

Moving on to multiterminal scenarios, the zero-delay constraint should be carefully defined. Specifically, suppose we have several non-cooperating encoders and one decoder. How does the decoder receive the bit-streams from each of the encoders and can it decode the message from one of the users before it starts decoding the other message? If the decoder must decode the messages simultaneously, then each encoder should use an independent instantaneous code which can depend only on the past. However, if for example, one of the messages can be decoded first (say, the messages are interleaved in one bit-stream), the first decoded message, as well as all past messages, can serve as SI when decoding the other message. This idea can be generalized to breaking the encoders messages into small pieces and sending them according to a predefined protocol.  We revisit the well known ``multiterminal source coding problem'' (two user Wyner-Ziv problem), where a source emits correlated pairs and each element of the pairs is observed by a different encoder. Each encoder sends a codeword to a joint decoder which reconstructs the current pair. The distortion between the reconstructions and the original variables should not exceed a given threshold. The challenge here is to find the set of achievable rates and distortions. This problem remained an open problem for over three decades (relevant literature review for this problem will be given in Section \ref{Sec:Multiterminal}). We consider a zero-delay version of this problem, where both users and the decoder must operate with zero-delay, assuming that one of the users messages is decoded first and serves as SI to the other user (simultaneous decoding of both messages is a simpler problem and its solution follow immediately from the solution to our problem). We show that, unlike the arbitrary delay setting, the zero-delay problem can be readily solved and the rate distortion region can be characterized. 


The remainder of this paper is organized as follows. In Section \ref{Sec:Prelims}, we give the formal setting and notation used throughout the paper. Section \ref{Sec:SingleUser} deals with single user problems. Multiterminal zero-delay source coding is handled in Section \ref{Sec:Multiterminal}. We conclude this work in Section \ref{Sec:Conclusion}.

\section{Preliminaries \label{Sec:Prelims}}
We begin with notation conventions. Capital letters represent scalar random variables (RV's), specific
realizations of them are denoted by the corresponding lower case letters, and their alphabet -- by calligraphic letters. For a positive integers $i,j$, $x^i_j$ will denote the vector $(x_j,\ldots, x_i)$. If $j=1$, the subscript will be omitted. The probability distribution over a finite alphabet $\calX$ will be denoted by $P_X(\cdot)$.  When there is no room for ambiguity, we will use $P(x)$ instead of $P_X(x)$.

For a joint distribution $P(x,y)$, we say that $x,x' \in \calX$ are \textit{confusable} if there is a $y\in\calY$ such that $P(x,y)>0$ and $P(x',y)>0$. A characteristic graph $G$ is defined on the vertex set of $\calX$ and $x,x'\in\calX$ are connected by an edge if they are confusable. The pair $(G,P)$, denotes a probabilistic graph consisting of $G$ together with the distribution $P$ over its vertices (here $P$ denotes the marginal on $\calX$). We say that two vertices $(x,x')$ are adjacent if there is an edge that connects them in $G$. The chromatic number of $G$, $\chi(G)$, is defined to be the smallest number of colors needed to color the vertices of $G$ so that no two adjacent vertices share the same color.

\begin{figure}[htp]
\centering
\includegraphics[width=0.9\textwidth]{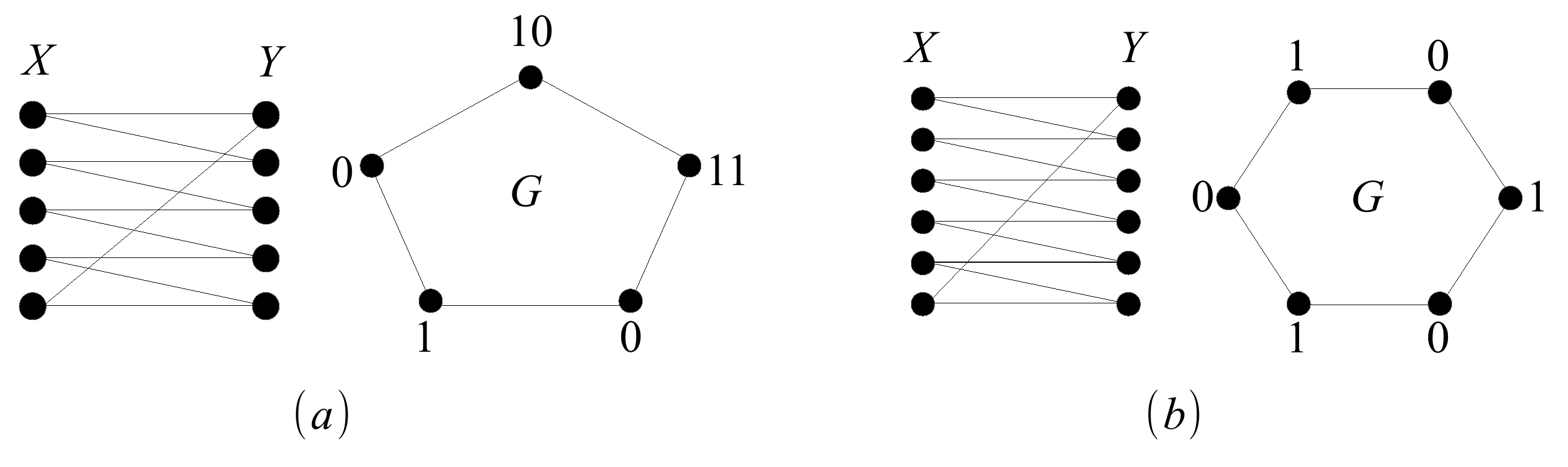} 
\caption{Example of bipartite graphs of $P(x,y)$ along with their associated characteristic graphs $G$ and a RI protocol for 5 (a) and 6 (b) letter alphabets with ``typewriter''  SI. \label{Fig:Pentagon}}
\end{figure}

We focus only on $(x,y)$ pairs with $P(x,y)>0$ and thus restrict attention only to \textit{restricted inputs} (RI) protocols, as defined in \cite{AlonOrlitski96}. A protocol for transmitting $X$ when the decoder knows $Y$, henceforth referred to as an RI protocol, is defined to be a mapping $\phi: \calX\to\cbr{0,1}^*$ such that if $x$ and $x'$ are confusable then $\phi(x)$ is neither equal to, nor a prefix of $\phi(x')$. An encoder that uses an RI protocol will be referred to as a SI-aware encoder. The length in bits of $\phi(x)$ will be denoted by $|\phi(x)|$.  Note that for restricted inputs, the prefix condition should be kept only over edges of $G$. Namely, for every $y\in\calY$, the prefix condition should be kept over the subset $\cbr{x: P(x,y)>0}$. The fact that the same $x\in\calX$ can be contained in multiple such subsets, but can have only a single bit representation, complicates the search for the optimal RI protocol.  
Let $\overline{l}_Y(\phi)\eqd \sum_{x\in\calX}p(x)|\phi(x)|$, where the subscript emphasizes that Y is known to the decoder. 
Let 
\begin{align}
	L_Y(X) = \min\cbr{\overline{l}_y(\phi) : \text{ $\phi$ is an RI protocol}}.  
\end{align}
In degenerate cases, where given a SI symbol there is only one possible source symbol with positive probability, no bits are needed to be sent and we set $|\phi(x)|=0$. In this case, the decoder knows that the next message in the bit-stream will be that of the next source symbol, so synchronization will be maintained.
Upper and lower bounds on $L_Y(X)$, in terms of the entropy of the optimal coloring, are given in \cite{AlonOrlitski96}.  There is no known closed form expression  for $L_Y(X)$. We will use $L_Y(X)$ as a figure of merit and our results will be single-letter expressions in terms $L_Y(\cdot)$. 
In Fig.\ref{Fig:Pentagon}, we give an example of bipartite graphs, formed by two joint distributions $P(x,y)$ where an edge connects $(x,y)$ if $P(x,y)>0$ along with the characteristic graphs and the optimal RI protocols for a uniform $P_X(\cdot)$.   


In Fig.\ref{Fig:Pentagon}a, we used 4 different bit representations for the source symbols. These bit representations are not prefix-free, but are easily seen to be uniquely decodable with the SI. The optimal bit representations imply a 4-coloring scheme for $G$, although $\chi(G)=3$. In Fig.\ref{Fig:Pentagon}b, however, $\chi(G)=2$ and indeed the optimal RI protocol uses a 2-coloring scheme. In Section \ref{Sec:Examples}, we will return to this example, as well as another example where increasing the quantizer output alphabet \textit{reduces} the rate.  

When the graph $G$ is complete, the prefix condition should be kept for all $x\in\calX$, thus reducing the RI protocol to regular prefix coding. In this case, $L_Y(X)$ is equal to the  average Huffman codeword length of $X$. This is, of course, also true when SI is not available at the decoder and we can think of the of a bipartite graph where all elements of $\calX$ are connected to the same constant $y$. The average Huffman length of the source $X$ will be denoted by $L(X)$.

In the proof the converses of our theorems, we use a ``genie'' that reveals common information to both encoder and decoder, thus we  define a conditional RI protocol. Let the triplet $(X,Y,Z)$ be distributed with some joint distribution $P(x,y,z)$. The information that is known to both parties will be denoted by $Z$, while $X$,$Y$ continue to play the roles of source output and the SI respectively. For any $z\in\calZ$, let $\Phi(z)$ denote the set of conditional RI protocols for $z$. Namely, the set of all RI protocols for source $X$ and SI $Y$ such that $p(x,y,z)>0$. For any $\phi\in\Phi(z)$, let $L_Y(\phi | z)\eqd \sum_{x\in\calX}P(x|z)|\phi(x)|$ be the average length when $Z=z$. Similarly, let 
\begin{align}
	L_Y(X|Z=z) = \min\cbr{\overline{l}_Y(\phi|z) : \phi\in\Phi(z)}.  
\end{align}
Finally, let $L_Y(X|Z) = \bE L_Y(X|Z=z)$ where the expectation is with respect to $P_Z(\cdot)$ and we used the same abuse of notation which is commonly used with the notation of conditional entropy. It follows that $L_Y(X|Z) \leq L_Y(X)$ since the set of RI protocols which are valid without the common knowledge of  $Z$ is contained in the set of conditional RI protocols which are valid when $Z$ is known at both ends.
In the special case where $Z=Y$, i.e., the SI is known to both parties, the RI protocol for each $y$ reduces to designing a Huffman code according to $P_{X|Y}(\cdot|y)$ for every $y\in\calY$. We will denote conditional Huffman length by $L(X|Z)$ and it is given by
\begin{align}
	L(X|Z) = \sum_{z} P(z) \min_{l(\cdot)}\sum_x P(x|z)l(x)
\end{align}
where in the minimization we consider all functions $l:\calX\to\mathbbm{R}^+$ that satisfy Kraft's inequality.

\section{Single-User Zero-Delay Problems}\label{Sec:SingleUser}
\begin{figure}[htp]
\centering
\includegraphics[width=0.9\textwidth]{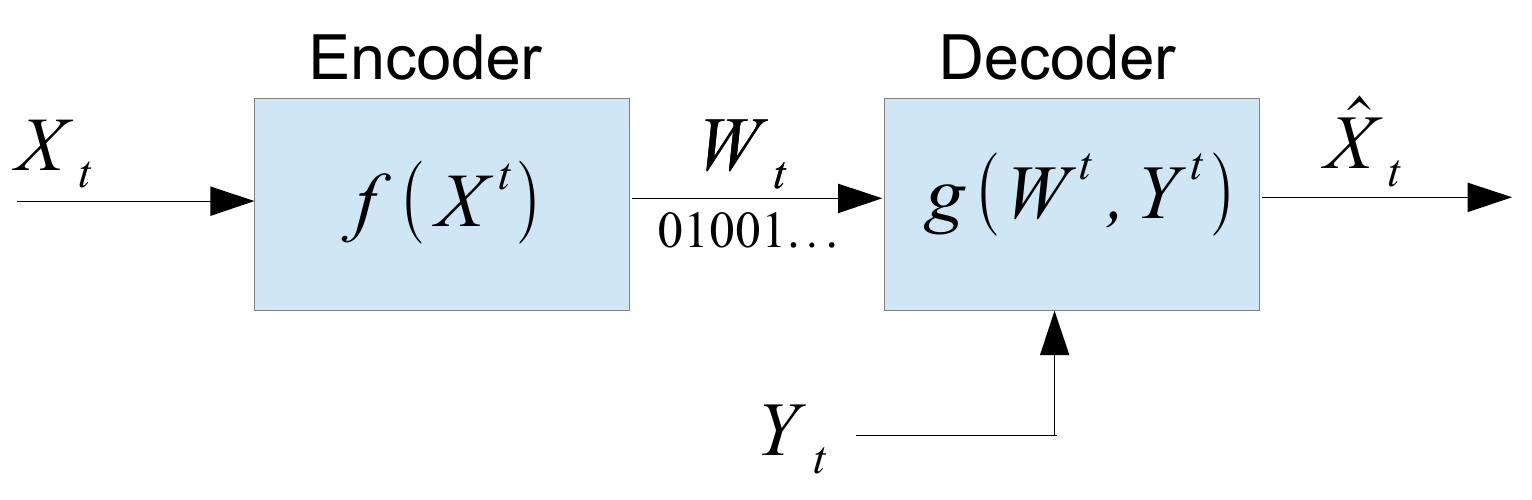} 
\caption{Basic model}\label{Fig:Basic_Model}
\end{figure}

We investigate the following zero-delay problem. A DMS is emitting pairs of random variables $(X_t,Y_t)$ according to $P(x,y)$, The alphabets of $X_t$ and $Y_t$, $\calX$ and $\calY$, as well as all alphabets in the sequel are finite. At time $t$, an encoder observes $X_t$ and transmits a compressed codeword, $W_t$, to a decoder which also observes $Y_t$. The decoder produces $\hX_t\in \hat{\calX}$, a reproduction of $X_t$, where $\hat{\calX}$ is the reproduction alphabet. Given a constant $D$ and a distortion measure $d:\calX\times\hat{\calX}\to \mathbbm{R}^+$, it is required that $\limsup_{n\to\infty}\frn\bE\sum_{t=1}^nd(X_t,\hX_t)\leq D$. Operation is in real-time. This means that the transmitted data, $W_t$, can be a function only of the encoder's observations no later than time $t$, namely, $X^t$. Similarly, the decoder's estimate, $\hX_t$ is a function of $(W^t,Y^t)$. Let $L_n$ denote the total number of bits sent after observing $n$ source symbols. The rate of the encoder is defined by $R\eqd \limsup_{n\to\infty} \frn\bE L_n$. Our goal is to find the tradeoffs between $R$ and $D$. Since no delay is allowed, $W_t$ must be encoded by an instantaneous code. The model is depicted in Fig.\ref{Fig:Basic_Model}.

In most block coding schemes, there is a probability of failure of the coding system (for example, the probability that the input will not be typical in schemes based on typicality). In these schemes, this probability becomes small as the block length increases and thus effectively does not affect the average distortion. In the zero-delay regime, however, there is no equivalent to failure since the encoding is sequential. An error at time $t$ will mean the decoder failed to parse the bit-stream sent by the encoder correctly. However, such an error event is catastrophic and will render the encoder future messages either meaningless or ambiguous. We therefore restrict our model to one where synchronization is maintained between the encoder and decoder at all times. This means that the encoder messages, $W_t$,  are sent without error. Note that in general $P(w_t,y_t)\neq P(w_t)P(y_t)$ (since $(X_t,Y_t)$ are not independent) and therefore, we will need to consider instantaneous coding of $W_t$ in the presence of correlated SI at the decoder. 

We start with a fully zero-delay system and give the rate-distortion function for this system. Later, we relax the zero-delay constraint on the SI and allow lookahead, showing that there is no performance gain. Furthermore we give the optimal performance when one of the components (encoder or decoder) is forced to be scalar, while no constraints are imposed on the other components (decoder or encoder). Subsection \ref{Sec:MainRes} summarizes the main results and discusses the results of this section. Proofs of Theorem \ref{Thm:RT_Y_LH} and \ref{Thm:ScalarDec} are given in Subsections \ref{Sec:P2PProof}  and \ref{Sec:ScalarProof}, respectively. Examples of SI aware quantizers that result from these theorems are given in Subsection \ref{Sec:Examples}. Finally, in Subsection \ref{Sec:CausalSet}, we revisit the causal setting as treated in \cite{NeuhoffGilbert1982} and \cite{TsachyNeri2005} and derive the equivalent of Theorem \ref{Thm:RT_Y_LH} for the causal setting, which was left open in \cite{TsachyNeri2005}.

\subsection{Main Results}\label{Sec:MainRes}

A pair $(R,D)$ is said to be achievable if there exists a rate-$R$ encoder with causal encoding functions $W_t=f_t(X^t), t=1,2,\ldots$, and a decoder with causal reproduction functions, $\hX_t=g_t(W^t,Y^t)$, such that the average distortion is smaller than $D$. Let $\calR_{ZD}(D)$ denote the infimum over all rates that are achievable with a given $D$, where the subscript stands for zero-delay. Let 
\begin{align}
	R_{ZD}(D) = \min_{h,f} L_Y(f(X)) \label{eq:RT_Def}
\end{align}
where the minimization is over all deterministic functions $h:\calZ \times \calY \to \hat{\calX}$ and 
$f:\calX\to\calZ$ such that $\bE d(X,h(Y,Z))\leq D$ (obviously, $|\calZ|\leq|\calX|$). Finally, denote the lower convex envelope of $R_{ZD}(D)$ by $\underline{R}_{ZD}(D)$. 

The first result of this paper is the following theorem:
\begin{theorem}\label{Thm:RTRD}
\begin{align}
 	\calR_{ZD}(D) = \underline{R}_{ZD}(D).
\end{align}
\end{theorem}
This theorem is proved along with Theorem \ref{Thm:RT_Y_LH} in Subsection \ref{Sec:P2PProof}. 

\noindent\textit{Remarks:\\}
\textbf{1}. Theorem \ref{Thm:RTRD} implies that optimal performance is attained by time-sharing at most two scalar SI-aware quantizers along with scalar decoders. The role of the function $f$ is to partition the source alphabet into subsets. Note that there is no sense in creating overlapping subsets since it will only increase the uncertainty at the decoder (increase the distortion) while adding edges to the characteristic graph of $f(X)$ with $Y$ (thus increasing the rate). 
For discrete $X$, there is a finite number of such partitions. For every possible partition, every choice of a possible $h$ will define a point in the $R-D$ plane. This point will be calculated by finding the best RI-protocol for the partition defined by $f$ and calculating the average distortion with $(f,h)$. 
Since there is a finite number of such $f,h$, there are a finite number of points on the $R-D$ plane. The lower convex envalope of these points gives us $\underline{R}_{ZD}(D)$, which will be piecewise linear. 

\textbf{2}. For a given $D$, the partition of the source alphabet with the minimal number of subsets which achieves distortion $D$ is not necessarily optimal and increasing the number of subsets (thus enlarging the quantizer's alphabet size) can \textit{reduce} the average rate.  To see this, assume that we have found $f,h$ where $f$ partitions the alphabet into the minimal number of subsets for this specific $D$. If at least one of these subsets contains more than one element, we can always find $f'$ and $h'$ such that $f'$ will divide this subset into two disjoint subsets, while $h'$ will be defined the same as $h$ for the original subset and therefore $f',h'$ will have the same average distortion as the original $f,h$. While it seems intuitive that the rate needed for $f'$ will be higher that the rate needed for $f$, this is not necessarily true. The reason is that the characteristic graph of $f(X)$ is fully connected, while this is not always the case for the characteristic graph of $f'(X)$.  To see why the characteristic graph of $f(X)$ is fully connected, note that if two subsets are not connected in the characteristic graph of $f(X)$, then they are not confusable. Therefore, we can combine these subsets into one subset without loss of performance since $h$ can act differently for the $Y's$ associated with the original subsets. But this contradicts the assumption that $f$ was the partition with the minimal number of subsets for the given $D$. If the characteristic graph of $f'(X)$ is not fully connected, the SI can be used to alleviate the prefix conditions on the code for $f'(X)$ and shorten its length. Therefore, increasing the output size of the quantizer might reduce the average length. We demonstrate such a phenomenon in Section \ref{Sec:ExampleNotFullyConnected}.

\textbf{3}. There is no loss of generality in the restriction to deterministic encoders since $L_Y(Z)$ is a concave functional of $\cbr{P(z|x)}$ while the distortion is linear $\cbr{P(z|x)}$. Therefore, optimizing over the whole convex set of stochastic encoders (represented by distributions $\cbr{P(z|x)}$) is equivalent to optimizing only over the extreme points of this set, which are the deterministic encoders. 

\textbf{4.} The result of this theorem relies heavily on the fact that no delay/ encoder lookahead is allowed. If, for example, the encoder will be allowed a lookahead of one symbol, i.e., $W_t = f(X^{t+1})$, then one can think of the problem as a zero-delay coding of a Markov source, with symbols $\tilde{X}_t=(X_t,X_{t+1})$. The results presented in \cite{Me2011} suggest that the optimal encoder in this case should use $\tilde{X_t}$ \textit{and} its previous messages\footnote{Although the result of \cite{Me2011} holds for a cost function which is a linear combination of the distortion and the average length, it can be adopted to the constrained setting as well when assuming that there is common randomness between encoder and decoder, see also \cite{Yuksel2007} for an example.}. While \cite{Me2011} does not rule out the existence of simpler optimal encoders, it seems intuitive that for Markov sources the encoder should not be scalar. 

Let $\calR_{ZD}^{\by}(D)$ denote the infimum over all rates that are achievable with a given $D$, with the same encoders as before and decoders that can use the whole SI sequence, i.e., $\hX_t = g_t(W^t,Y^n)$. We have the following theorem:
\begin{theorem} \label{Thm:RT_Y_LH}
\begin{align}
	\calR_{ZD}^{\by}(D) = \underline{R}_{ZD}(D).  
\end{align}
\end{theorem}
We prove Theorem \ref{Thm:RT_Y_LH} in Section \ref{Sec:P2PProof}. Theorem \ref{Thm:RTRD} is a special case of Theorem \ref{Thm:RT_Y_LH} and the direct part of the proof, as apparent from the theorems, is the same for both theorems. 

The  theorem states that allowing the decoder to observe the future SI symbols will not result in any performance gain. This is in contrast to the setting of non-causal access to the source and causal access to the SI at the decoder, treated in \cite{TsachyElGamal06}, where it was shown that SI look-ahead can improve performance.

Now assume that the encoder, in addition to the source $X_t$, observes causally another process, $S_t\in\calS$, which is jointly distributed with the source and SI according to some joint distribution $P(S^n,X^n,Y^n)= \prod_{t=1}^n P(S_t,X_t,Y_t)$, i.e., the encoder has access to SI as well as the source. The goal remains to reconstruct only the source $X_t$ with distortion not exceeding $D$. Let us define $\calR_{ZD, EncSI}(D)$ to be the infimum over all rates that are achievable with a given $D$, where the encoders are causal functions  $W_t=f_t(S^t,X^t)$ and the decoders are as in Theorem \ref{Thm:RT_Y_LH}, i.e., causal in $\cbr{W_t}$ and access the whole SI. Let us define 
\begin{align}
	R_{ZD, EncSI}(D)= \min_{h,f} L_Y(f(X,S))
\end{align}
where the minimization is over all deterministic functions $h:\calZ \times \calY \to \hat{\calX}$ and 
$f:\calX\times\calS\to\calZ$ such that $\bE d(X,h(Y,Z))\leq D$. We have the following corollary which follows from Theorem  \ref{Thm:RT_Y_LH}, and is proved exactly in the same way
\begin{corrolary}
\begin{align}
	 \calR_{ZD, EncSI}(D) = R_{ZD, EncSI}(D). 
\end{align}

\end{corrolary}
Unlike the case without the SI, where causal encoder SI cannot help (since it can be simulated by the encoder even if it is not available), when the decoder has SI, encoder SI indeed helps and can be viewed as extra encoder insight on the decoder SI. In the extreme case when $S_t=Y_t$, we obtain back the results of \cite{TsachyNeri2005}, adopted to the zero-delay setting. Note that with $S$ at the encoder, $(x,x')$ will be confusable of there exist  $(y,s)$ such that $P(s,x,y)>0$ and $P(s,x',y)>0$ and thus, even if no distortion is allowed, $S$ can be used to reduce the number of edges in the characteristic graph $G$ and thus reduce the constraints on the RI protocol,  achieving a lower average rate.

From Theorems \ref{Thm:RTRD} and \ref{Thm:RT_Y_LH}, it is apparent that as long as the encoding function is causal in the source, and the decoder is causal in the encoder messages, scalar pairs of encoders and decoders (codecs) are optimal. This suggests, as stated in the Introduction, that the optimal performance achievable by zero-delay systems is by far inferior to the performance of systems that allow arbitrary delay. The inquisitive reader might ask at this point which of the three constraints imposed by our zero-delay model (causal encoding function / causal decoding functions / sequential messages with instantaneous code) is to be ``blamed'' for the rather poor performance  given by Theorems \ref{Thm:RTRD} and \ref{Thm:RT_Y_LH}. Had we alleviated one of these constraints, can optimal rate-distortion performance be achieved?

We now show by example that at least without SI at the decoder, there are sources and distortion measures for which the answer to the last question is affirmative. To see this, think of a system with an encoder that observes the whole sequence $X^n$. The encoder can send no more than $\log|\hat{\calX}|$ bits per transmission and the decoders are causal (sequential) in the encoder's messages, meaning that $\hX_t$ is calculated using the data received in the first $t$ encoder transmissions. Compared to the systems described above, we only alleviated the constraint on the encoder. Note that we have to restrict the number of bits the encoder can send in each transmission, otherwise, there is no meaning to the decoder being sequential since the encoder can send the description for $\hX^n$ in a single transmission, as done in block-coding schemes. Such a scheme can be extremely useful for streaming purposes, where the whole stream is known in advance to the encoder and it should be streamed to a decoder without decoding delay. Our scheme works as follows: A classical rate-distortion random code \cite{cover} is used. The encoder finds the first codeword ($\hX^n$) in the codebook which is distortion-typical (as defined in \cite{cover}) and starts to transmit the \textit{reproduction symbols}, $\hX_t$, sequentially, using $\log|\hat{\calX}|$ bits each transmission (in contrast to the classical encoder which will send the index of the codeword in a block). The decoder outputs $\hX_t$ as it receives it. 
The idea here, is that after receiving enough symbols, the decoder can detect the specific codeword in the codebook since, with high probability, no other codeword will have the same prefix and the rest of the reproduction symbols can be reproduced without further transmissions from the encoder. We show in Appendix \ref{App:Scheme} that such a scheme can can achieve optimal rate distortion performance ($R(D)$ bits per source sample) in some cases and is less than one bit away from optimality in the remainder cases. Tree codes are another example of sequential schemes (see \cite{Merhav2010} and references therein), but there, the fact that the rate is constrained to be a the natural logarithm of an integer usually forces the schemes to work in small blocks, unlike the scheme presented above. The point we make here is that at least for some combinations of sources and distortion functions, causality constraints only on the decoders and the use of instantaneous codes does not affect the performance.

Next, we ask what is the best attainable performance if we are constrained to use a scalar decoder/encoder, when the other component (encoder/decoder) is unconstrained. For example, we have a scalar encoder, but the decoder can wait until it received $(W^n, Y^n)$ and only then output $\hX^n$ or vice-verse.  Such a limitation is motivated by practical constraints imposed on the encoding/decoding devices, for example, the encoder is a simple sensor but the decoder which receives data from another sensor as SI can be as complex as needed for optimal performance. Note that for a scalar decoder, although we allow a non-causal encoder, we still require that the encoder will send a message every time instance and the decoder will reconstruct $\hX_t$ according to this message and SI. 
Can the simplicity of one of the elements be compensated by the complexity of the other? The next theorem will assert that the answer to this question is negative.

Let $\calR_{s-d}(D)$ denote the infimum over all rates that are achievable with a given $D$, when non-causal encoders are allowed, i.e., $W_t=f_t(X^n)$, but the decoders are restricted to be scalar in $W_t$, i.e., $\hX_t = g_t(W_t,Y^n)$ (the s-d subscript stands for scalar decoder). Similarly, let $\calR_{s-e}(D)$ denote the infimum over all rates that are achievable with a given $D$, when non-causal decoders are allowed, i.e., $\hX_t=g_t(W^n,Y^n)$, but the encoders are restricted to be scalar in $X_t$, i.e., $W_t = f_t(X_t)$.
We have  the following theorem:
\begin{theorem}\label{Thm:ScalarDec}
\begin{align}
 	\calR_{s-d}(D) = \calR_{s-e}(D) = \underline{R}_{ZD}(D).  
\end{align}
\end{theorem}
Theorem \ref{Thm:ScalarDec} states that when either the encoder or decoder are constrained to be scalar, the other side can be scalar as well without any performance loss. Note that for the scalar decoder case, this is true even if the decoder is scalar only in the encoder's messages but has full SI look-ahead. This theorem extends Theorem 5 of \cite{GaarderSlepianS79} to include variable rate, look-ahead and SI. 

The next two subsections will be devoted to the proofs of Theorem \ref{Thm:RT_Y_LH} and the right hand side of Theorem \ref{Thm:ScalarDec}. The proof that $\calR_{s-d}(D)  = \underline{R}_{ZD}(D)$, which follows the same lines, is deferred to Appendix \ref{App:Scalar_Enc_Proof}. Examples are given in Subsection \ref{Sec:Examples}. Finally, the causal setting will be treated in Section \ref{Sec:CausalSet}.

\subsection{Proof of Theorem \ref{Thm:RT_Y_LH}}\label{Sec:P2PProof}
\subsubsection{Converse part}
At every stage, the encoder sends a message $W_t$. This message is in general a function of $X^t$. The decoder at time $t$ has already received $W^{t-1}$ and has access to $Y^n$. Only the current and past SI, $Y^t$, serves as SI when sending $W_t$ since $Y^{n}_{t+1}$ is independent of $X^t$. Therefore, we have
\begin{align}
 	nR \geq \sum_{t=1}^n L_{Y^t}(W_t|W^{t-1}).
\end{align}
To avoid the complex SI structure, which depends on the time instance $t$, we use a genie aided scheme. At each time instant, a genie reveals all past SI symbols to the encoder and all past source symbols to the decoder. With this ``genie aided" feedback and feed--forward, at each time instant, only $Y_t$ serves as SI which is not known to both parties. Therefore, the minimal average number of transmitted bits at each stage is lower bounded by $L_{Y_t}(W_t|X^{t-1},Y^{t-1})$. For any sequence of encoders which are functions of $(X^t,Y^{t-1})$ and any sequence of reproduction decoders which are functions of $(W_t,X^{t-1},Y^n)$, satisfying the distortion constraint, we have:
\begin{align}
 	nR &\geq \sum_{t=1}^n L_{Y_t}(W_t|X^{t-1},Y^{t-1})\nl
	&\geq \sum_{t=1}^n L_{Y_t}(W_t|X^{t-1},Y^{t-1}, Y_{t+1}^{n})\label{eq:RTConv_Cond}\\
	&=\sum_{t=1}^n \int L_{Y_t}(W_t|x^{t-1},y^{t-1},y_{t+1}^{n})d\mu(x^{t-1},y^{t-1},y_{t+1}^{n})\nl
 	&=\sum_{t=1}^n \int L_{Y_t}(f_t(X_t,x^{t-1},y^{t-1})|x^{t-1},y^{t-1},y_{t+1}^{n})d\mu(x^{t-1},y^{t-1},y_{t+1}^{n})\nl
	&=\sum_{t=1}^n \int L_{Y_t}(f_t(X_t,x^{t-1},y^{t-1}))d\mu(x^{t-1},y^{t-1},y_{t+1}^{n})\label{eq:RTConvPre}
\end{align}
where in \eqref{eq:RTConv_Cond} we used the fact that conditioning reduces the average length. In the next line, $\mu(\cdot)$ denotes the joint probability mass function of its arguments and the last equation is true since $X_t$ is independent of $(X^{t-1},Y^{t-1},Y_{t+1}^{n})$.
Now, $f_t(X_t,x^{t-1},y^{t-1})$ can be seen as a specific choice of $f(X_t)$ in the definition of $R_{ZD}(D)$, \eqref{eq:RT_Def}. This, along with the fact that we know that $Y^{t-1}=y^{t-1}, Y_{t+1}^{n}=y_{t+1}^{n}$ and $X^{t-1}=x^{t-1}$, makes the decoding function \\
$\hX_t = g_t(f_t(X_t,x^{t-1}),x^{t-1},y^{t-1},Y_t, y_{t+1}^{n})$ a specific choice of $h(\cdot,\cdot)$ in the definition of  $R_{ZD}(D)$. We therefore have
\begin{align}
 	&nR \geq \sum_{t=1}^n \int L_{Y_t}(f_t(X_t,x^{t-1},y^{t-1}))d\mu(x^{t-1},y^{t-1},y_{t+1}^{n})\nl
	&\geq \sum_{t=1}^n\int R_{ZD}(\bE [d(X_t, g_t(f_t(X_t,x^{t-1},y^{t-1}),x^{t-1},y^{t-1},Y_t),y_{t+1}^n)|x^{t-1},y^{t-1},y_{t+1}^n])\times\nl
	&~~~~~~d\mu(x^{t-1},y^{t-1},y_{t+1}^{n})\label{eq:RTConv1}\\
	&\geq \sum_{t=1}^n\int \underline{R}_{ZD}(\bE [d(X_t, g_t(f_t(X_t,x^{t-1},y^{t-1}),x^{t-1},y^{t-1},Y_t,y_{t+1}^{n}))|x^{t-1},y^{t-1},\nl
	&~~~~~~y_{t+1}^{n}])d\mu(x^{t-1},y^{t-1},y_{t+1}^{n})\label{eq:RTConv2}\\
	&\geq \sum_{t=1}^n \underline{R}_{ZD}\Bigg( \int\bE [d(X_t, g_t(f_t(X_t,x^{t-1},y^{t-1}),x^{t-1},y^{t-1},Y_t,y_{t+1}^{n}))|x^{t-1},y^{t-1},\nl
	&~~~~~~y_{t+1}^{n}]d\mu(x^{t-1},y^{t-1},y_{t+1}^{n})\Bigg)\label{eq:RTConv3}\\
	&= \sum_{t=1}^n\underline{R}_{ZD}\rbr{ \bE \sbr{d(X_t, g_t(f_t(X^{t},Y^{t-1}),X^{t-1},Y^{n}))}}\nl
	&= \sum_{t=1}^n \underline{R}_{ZD}\rbr{ \bE \sbr{d(X_t, \hX_t)}}\nl
	&\geq n \underline{R}_{ZD}\rbr{ \frn\sum_{t=1}^n\bE \sbr{d(X_t, \hX_t)}}\label{eq:RTConv4}\\
	&\geq n\underline{R}_{ZD}\rbr{D},\label{eq:RTConv5}
\end{align}
where \eqref{eq:RTConv1} follows from the definition of $R_{ZD}(D)$ and the discussion following \eqref{eq:RTConvPre}, \eqref{eq:RTConv2} follows from the definition of  $\underline{R}_{ZD}(D)$,  \eqref{eq:RTConv3} and \eqref{eq:RTConv4} follow from the convexity of $\underline{R}_{ZD}(D)$. Finally, \eqref{eq:RTConv5} follows from the monotonicity of $\underline{R}_{ZD}\rbr{D}$.
Combining the above with the direct part given in the next subsection, we also proved that feedback of the SI and feed-forward of the source cannot improve performance here. 

\subsubsection{Direct part}
The direct part of the theorem is obtained by time--sharing two scalar SI-aware quantizers.  By definition of $\underline{R}_{ZD}\rbr{D}$, we have that there exist $(D_1,D_2,\lambda)$ such that $D = \lambda D_1+(1-\lambda)D_2$ and $(f_1,h_1), (f_2,h_2)$ that are the achievers of $R_{ZD}\rbr{D_1}$ and $R_{ZD}\rbr{D_2}$, respectively, such that $\lambda R_{ZD}\rbr{D_1} + (1-\lambda)R_{ZD}\rbr{D_2} = \underline{R}_{ZD}\rbr{D}$. Let $\phi_1$,$\phi_1$ be the optimal protocols for $Z_{1,t}=f_1(X_t),Z_{2,t}=f_2(X_t)$ respectively. Also, let $k_n\leq n$ be a non-decreasing  sequence of integers such that 
$\lim_{n\to\infty}\frac {k_n} n = \lambda$. For every $n$, we use $(f_1,h_1)$ for the first $k_n$ stages and $(f_2,h_2)$ for the rest of the $n$--block.  The resulting $Z_{i,t}, i=1,2$, are coded with the optimal protocols $\phi_1$ or $\phi_2$. The average distortion of this scheme is given by
\begin{align}
 	&\frn\sum_{t=1}^n\bE d(X_t, g(Y^t,Z^t)) \nl
	&~~~~= \frac {k_n}{n}\bE d(X_t, h_1(Y_t,Z_{1,t}))+\frac {n-k_n} n \bE d(X_t, h_2(Y_t,Z_{2,t}))\nl
	&~~~~\leq\frac{k_n}{n}D_1 + \frac {n-k_n} {n} D_2 
\end{align}
and therefore, $\lim_{n\to\infty}\frn\sum_{t=1}^n\bE d(X_t, g(Y^t,Z^t)) \leq D$.
The rate of the code is given by
\begin{align}
 	\frn\bE L_n &=\frn\sum_{t=1}^{k_n}\bE|\phi_1(Z_{1,t})| +  \frn\sum_{t=k_n+1}^{n}\bE|\phi_2(Z_{2,t})|\nl
	&= \frac {k_n} n L(f_1(X)) + \frac{n-k_n} {n}L(f_2(X))
\end{align}
Therefore,
\begin{align}
	R &= \lim_{n\to\infty} \frn \bE L_n \nl
	&=\lambda L_Y(f_1(X)) + (1-\lambda)L_Y(f_2(X))\nl
	&= \underline{R}_{ZD}\rbr{D}.
\end{align}

\subsection{Proof of Theorem \ref{Thm:ScalarDec}}\label{Sec:ScalarProof}
In this subsection, we prove the first part of Theorem \ref{Thm:ScalarDec}, namely, we show that $\calR_{s-d}(D) = \underline{R}_{ZD}(D)  
$. The proof that $\calR_{s-e}(D) = \underline{R}_{ZD}(D)$ follows the same lines and is deferred to Appendix \ref{App:Scalar_Enc_Proof}. Since the direct part is the same as the direct part of Theorem \ref{Thm:RT_Y_LH}, we need to prove the converse only.

We use the same genie-aided scheme as in the proof of Theorem \ref {Thm:RT_Y_LH} and prove a stronger converse:  For any sequence of encoders, with possible access to the whole source sequence and past SI, $W_t=f_t(X^n, Y^{t-1})$ and any sequence of decoders with access to the whole SI sequence, past source symbols and the current encoder output, $\hX_t=g_t(W_t,X^{t-1},Y^n)$, that achieve average distortion $D$ (i.e., $\frn\sum_{t=1}^n\bE d(X_t,\hX_t)\leq D$), we have
\begin{align}
 	nR & \geq\sum_{t=1}^n \int L_{Y_t}(W_t|X^{t-1},Y^{t-1})\nl
	& \geq\sum_{t=1}^n \int L_{Y_t}(W_t|X^{t-1},X_{t+1}^n, Y^{t-1}, Y^{n}_{t+1})\nl
	&=\sum_{t=1}^n \int L_{Y_t}(W_t|x^{t-1},x_{t+1}^n, y^{t-1}, y_{t+1}^n)d\mu(x^{t-1},x_{t+1}^n, y^{t-1}, y_{t+1}^n)\nl
	&=\sum_{t=1}^n \int L_{Y_t}(f_t(x^{t-1},X_t,x_{t+1}^n, y^{t-1}) |x^{t-1},x_{t+1}^n, y^{t-1}, y_{t+1}^n)\times\nl
	&~~~~~~d\mu(x^{t-1},x_{t+1}^n, y^{t-1}, y_{t+1}^n)\nl
	&\geq\sum_{t=1}^n \int L_{Y_t}(f_t(x^{t-1},X_t,x_{t+1}^n, y^{t-1}))d\mu(x^{t-1},x_{t+1}^n, y^{t-1}, y_{t+1}^n)
\end{align}
where the last step, as before, relied on the memorylessness  of the source. Now, the rate in the inner expression cannot be smaller than the rate of the optimal scalar system (with all conditioned elements serving as index of functions), which achieves the distortion achieved by the given decoding function, $h_t(W_t,y^{t-1},Y_t,y_{t+1}^n)$. Therefore,
\begin{align}
	nR&\geq\sum_{t=1}^n \int L_{Y_t}(f_t(x^{t-1},X_t,x_{t+1}^n, y^{t-1}))d\mu(x^{t-1},x_{t+1}^n, y^{t-1}, y_{t+1}^n)\nl
	&\geq \sum_{t=1}^n \int \underline{R}_{ZD}\left(\bE \left[d(X_t, g_t(f_t(x^{t-1},X_t,x_{t+1}^n, y^{t-1}),x^{t-1}, y^{t-1},Y_t,y_{t+1}^n))\right.\right.\nl
	&\left.\left.~~~~|x^{t-1},x_{t+1}^n, y^{t-1}, y_{t+1}^n\right]\right)d\mu(x^{t-1},x_{t+1}^n, y^{t-1}, y_{t+1}^n)\nl
	&\geq \sum_{t=1}^n  \underline{R}_{ZD}\Bigg(\int \bE \left[d(X_t, g_t(f_t(x^{t-1},X_t,x_{t+1}^n, y^{t-1}),x^{t-1}, y^{t-1},Y_t,y_{t+1}^n))\right. \nl
	&\left.~~~~|x^{t-1},x_{t+1}^n, y^{t-1}, y_{t+1}^n\right]d\mu(x^{t-1},x_{t+1}^n, y^{t-1}, y_{t+1}^n)\Bigg)\label{eq:scalarConv1}\\
	&\geq \sum_{t=1}^n  \underline{R}_{ZD}\rbr{\bE \sbr{d(X_t, g_t(f_t(X^n),X^{t-1}, Y^n)}}\nl
	&\geq \underline{R}_{ZD}\rbr{ \sum_{t=1}^n\bE \sbr{d(X_t, \hX_t)}}\label{eq:scalarConv2}\\
	&\geq \underline{R}_{ZD}\rbr{D}\label{eq:scalarConv3}
\end{align}
where in \eqref{eq:scalarConv1} and \eqref{eq:scalarConv2} we used the convexity of $\underline{R}_{ZD}(\cdot)$ and its monotonicity in \eqref{eq:scalarConv3}.

\subsection{Examples}\label{Sec:Examples}
\subsubsection{Lossless transmission}
It is interesting to relate the above results to the lossless case which was considered in \cite{AlonOrlitski96}. Since $D=0$ cannot be achieved by time-sharing positive distortions, we get that 
\begin{align}
	L_Y(X) = \min_{h, f: h(y,f(x))=x}L_Y(f(X)). \label{eq:ZeroD} 
\end{align}
Let $Z=f(X)$. Any $f$ which is a coloring of the characteristic graph $G$ and $h$, which is the mapping from color and $y$ back to $x$, are valid candidates in the optimization problem of \eqref{eq:ZeroD}. If $f$ is not a valid coloring, meaning that two connected $x_1,x_2$ will result in the same $z$, then there is no $h$ which can result in zero error. In essence, we are looking for the coloring for which the restricted inputs protocol will produce the smallest rate. Note that when searching for the best coloring, our performance will be affected only by the characteristic graph $G_z$ which will be built with the ``source'' $(Z,Y)$. If $f$ is a minimal coloring, i.e, $z\in\cbr{1,2,\ldots,\chi(G_z)}$, then $G_z$ is complete. To see this, note that for a minimal coloring, if $z_1$ and $z_2$ are not connected, then these colors can be combined and this reduces the number of colors, contradicting the fact that $f(x)$ is a minimal coloring. Remember that a complete graph reduces the RI protocol to Huffman coding. This means that the SI is not helping us to code the colors. Therefore, looking for colorings which will induce an incomplete $G_z$ (i.e., non-optimal coloring) will allow us to use the SI not only to reduce the alphabet of the encoder output, but also for the coding of its output (namely, relax the prefix condition on the codewords when the graph is complete). In the example of Fig.\ref{Fig:Pentagon}a, we used a 4-coloring scheme (we had 4 different bit representations for the vertices of $G$) and not the optimal 3-coloring. Indeed, $G_z$ for the 4-coloring is not complete. For a uniform source, we get an average rate of 1.4 with the 4 coloring and if we had used a 3-coloring we would get an average rate of 1.6.  

\subsubsection{Uniform source, fully connected SI model}
Let the reconstruction alphabet be the same as the source alphabet. We use the Hamming distortion measure ($d(x,\hx) =0$ if $x=\hx$ and $d(x,\hx) =1$ otherwise). The encoder partitions the source alphabet into disjoint subsets $\calA_1,\calA_2,\ldots,\calA_k$, $k\leq |\calX|$. When the encoder observes a new source symbol, $x$, it sends the index of the subset containing $x$, using an RI protocol, as defined in Section \ref{Sec:Prelims}. With the Hamming distortion measure, the average distortion is equal to the probability of error. Therefore, the optimal decoder is the maximum likelihood decoder, namely:
\begin{align*}
	\hx = \argmax{x} P(y,z|x) = \argmax{x\in\calA_z}P(y|x).	
\end{align*}
where $z$ is the subset index, sent by the encoder. 

Let $|\calX|=|\calY|=M$ and let for a small constant $p$, $P(X=a)=\frac 1 M$, $P(y=\alpha|x=a)=1-p$ if $\alpha=a$ and $P(y=\alpha|x=a)=\frac p {M-1}$ for any $\alpha\neq a$. With this choice of the joint distribution, since the bipartite graph of $\cbr{P(x,y)}$ is fully connected, then so is the bipartite graph of $\cbr{P(y,z)}$, regardless of the choice of Z. Therefore, the RI protocol used to describe the index of the subsets is reduced to a Huffman code for $Z$. 

It is shown in \cite{ReaniMerhav2012}, that for this distortion measure and source, only the number of partitions, and not their content (i.e., the actual alphabet letters in each subset), affects the average distortion. The distortion as a function of the number of partitions, $K$, is given by $\frac p {M-1}(|\calX|-K)$. It turns out that in this case, $\underline{R}_{ZD}\rbr{D}=L(X)-\frac {L(X)} {p}D$, which is obtained by time-sharing the two trivial quantizers: the one that does not send information $(R=0,D=p)$ and the lossless quantizer $(R=L(X),D=0)$. 

\subsubsection{Uniform source, given SI model}\label{Sec:ExampleNotFullyConnected}
\begin{figure}[htp]
\centering
\includegraphics[width=0.7\textwidth]{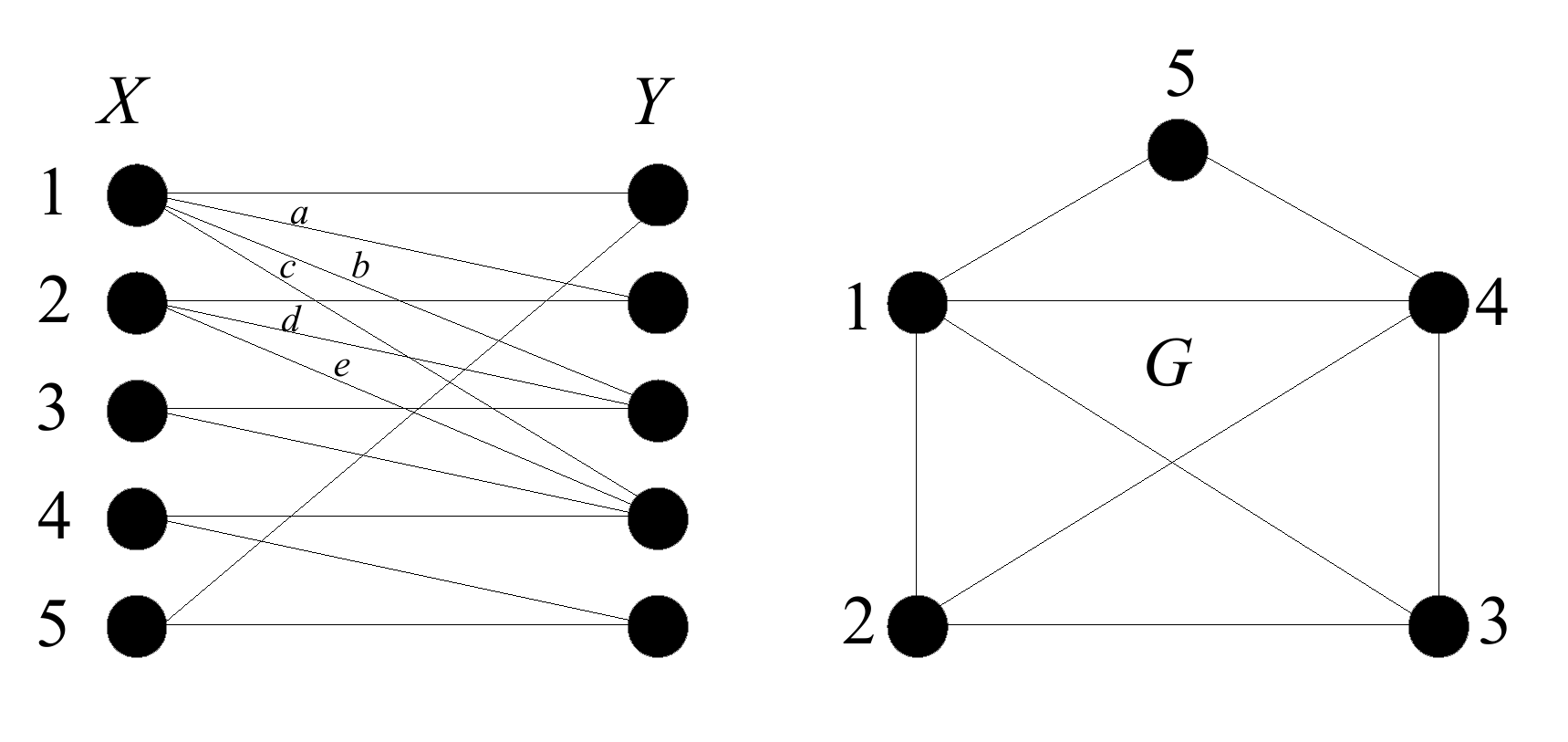} 
\caption{$P(y|x)$ and the resulting characteristic graph.}\label{Fig:Example}
\end{figure}

\begin{figure}[htp]
\centering
\includegraphics[width=0.7\textwidth]{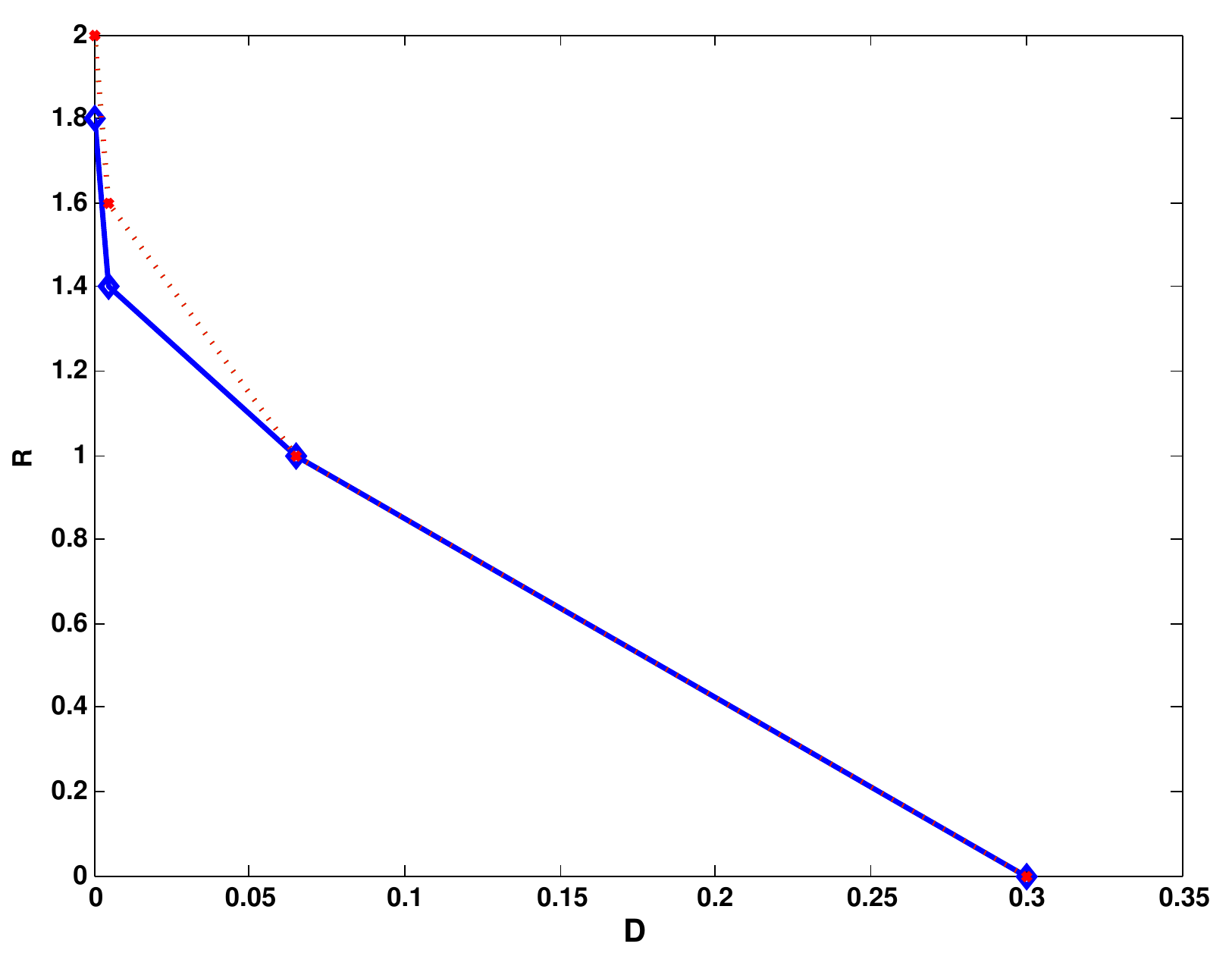} 
\caption{$\underline{R}_{ZD}\rbr{D}$ (solid) compared to a system that uses the SI only for reproduction (dotted) with $p=0.3$.}\label{Fig:ExamplePlot}
\end{figure}

We continue with the Hamming distortion measure and a uniform source with $|\calX|=5$. The channel from $X$ to $Y$ is given in Fig.\ref{Fig:Example}, along with the characteristic graph of $X$.
In this example we set for $\alpha\in\calX$: $P(y=\alpha|x=\alpha)=1-p$ and $a=\half p,b=\frac 5 {12} p, c=\frac 1 {12} p, d=\frac 3 4 p,e=\frac 1 4 p$. Note that the chromatic number of $G$ is 4. This means that any partitioning of the alphabet of $X$ into less than 4 subsets will incur a lossy reconstruction. Unlike the previous example where the SI could be used only in the reconstruction, but not to reduce the length of the transmission (since $G$ was fully connected), here it will be used for both. Note that as in the example of Fig.\ref{Fig:Pentagon}, the optimal rate for lossless transmission is actually obtained by using more subsets than the chromatic number of $X$. The optimal two subset partition ($|\calZ|=2$) is $\cbr{1,4},\cbr{2,3,5}$, yielding an average distortion of $\frac {13} {60}p$. The rate for this (and any binary) partition is $1$. The optimal 3-subset partition is $\cbr{1,4},\cbr{2,5}, \cbr{3}$ yielding an average distortion of $\frac 1 {60} p$. The average rate for this partitioning is $8/5$. However, in this case it is beneficial to split $\cbr{2,5}$ and obtain a lower rate of $7/5$ (using the SI to alleviate the prefix requirement). Although we use more subsets, the rate is reduced, as discussed in the remarks that followed the statement of Theorem \ref{Thm:RTRD}. In Fig.\ref{Fig:ExamplePlot}, we compare $\underline{R}_{ZD}\rbr{D}$ to the performance of a system that uses the minimal number of subsets for each $D$ and uses Huffman codes, i.e., a system that uses the SI only for the reproduction but not for the compression. 

\subsection{The Causal Setting}\label{Sec:CausalSet}
In this section, we revisit the causal setting considered in \cite{NeuhoffGilbert1982} and \cite{TsachyNeri2005}. In this setting, we still restrict the encoding and decoding functions to be causal functions of the their respective inputs, however, we do not restrict the delay which is imposed by the system. Effectively, this means that we drop the instantaneous code constraint and allow more sophisticated coding techniques which introduce delays.  In \cite{NeuhoffGilbert1982}, \cite{TsachyNeri2005}, the SI was either not present or was not used in the reconstruction of $\hX_t$ (unless it was available at both the encoder and decoder) 
and therefore, the encoder could calculate the reproduction symbols. This in turn, enabled to recast the system into a reproduction coder (which creates $\cbr{\hX_t}$) followed by a lossless entropy coder, which coded $\hX^n$ into a message that was sent to the decoder (this is where the delay was introduced). However, when the SI is available only at the decoder and used in the reconstruction, the encoder cannot reproduce $\cbr{\hX_t}$ and therefore the encoder and the reconstruction decoder must be decoupled. In \cite{TsachyNeri2005}, the model we treat here was left unsolved. 
The combination of encoding that introduces delay with causal decoding functions is less motivated in practical terms than the zero-delay models that were introduced in the previous subsections. However, it can represent situations where the lossless coding of the encoder messages is done separately and then streamed to a simple causal decoder. The results which show that scalar codecs are optimal support such a model in retrospect. 

In contrast to the zero-delay model, where synchronization must be kept between encoder and decoder over the bit-stream and therefore no decoding errors are allowed, here, since block coding is allowed, we have no such constraint. A block error with a vanishing probability will translate into a vanishing additional distortion as with the classical rate-distortion solutions. 

Let $\calR_c(D)$ denote the infimum over all rates such that there exist causal encoders $W_t=f_t(X^t)$, decoders which are causal in the encoder's messages $\hX_t=g_t(W^t,Y^n)$ and define
\begin{align}
 	R_c(D) = \inf_{f,h}H(f(X)|Y) \label{eq:CausalRD}
\end{align}
where the minimum is over all functions $f:\calX\to\calZ$ and $h: \calZ\times\calY\to\hat{\calX}$ such that $\bE d(X,h(f(X),Y))\leq D$. Finally, define  $\underline{R}_c(D)$ to be the lower convex envalope of $R_c(D)$. 
We have the following theorem:
\begin{theorem}
\begin{align}
 	\calR_c(D) = \underline{R}_c(D).
\end{align}
\end{theorem}

We will only outline the proof since after a few steps it is similar to the proofs of the previous theorems when replacing the average length functional with the entropy functional. In the direct part, we again timeshare at most two scalar encoders and then code the resulting blocks using a Slepian-Wolf  \cite{SlepianWolf73} code. For the converse, we start with the converse to the lossless source coding with SI, stating that $nR\geq H(W^n|Y^n)$. While this step is valid when the SI in known at both sides when losslessly coding $W^n$, and might seem a loose lower bound here, we know from the Slepian-Wolf theorem that there is essentially no loss when the SI is available only at the decoder and a vanishing probability of block error is allowed. We have:
\begin{align}
	nR &\geq H(W^n|Y^n)\nl
	&= \sum_{t=1}^n H(W_t|W^{t-1},Y^n)\nl
	&\geq \sum_{t=1}^n H(W_t|W^{t-1},X^{t-1},Y^n)\nl
	&= \sum_{t=1}^n H(W_t|X^{t-1},Y^n)\nl
	&= \sum_{t=1}^n \int H(f_t(X_t, x^{t-1})|x^{t-1},y^{t-1},Y_t,y_{t+1}^n)d\mu(x^{t-1},y^{t-1},y_{t+1}^n)\nl
	&= \sum_{t=1}^n \int H(f_t(X_t, x^{t-1})|Y_t)d\mu(x^{t-1},y^{t-1},y_{t+1}^n)
\end{align}
 From here, one can identify that $R_c(D)$ can further lower bound the conditional entropy in the last equation. The rest of of the proof essentially follows the steps of the proof of Theorem \ref{Thm:RT_Y_LH} after equation \eqref{eq:RTConv1}.

\section{Multiterminal Problem}\label{Sec:Multiterminal}
In this section we revisit the multiterminal source coding problem (two-user Wyner-Ziv problem), where two correlated random variables are observed separately by two non-cooperative encoders who communicate with a decoder. The decoder needs to reconstruct both sources and the distortion between the reconstructions and the original variables should not exceed a given threshold. As mentioned in the Introduction, this problem has been open for the general setting for over three decades. There are specific cases in which the arbitrary delay problem can be solved for a general source and distortion measure. When no distortion is allowed, this is the Slepian-Wolf problem \cite{SlepianWolf73}  and when one of the variables is known to the decoder, this is the original Wyner-Ziv problem \cite{WynerZiv76}. Other examples include the source coding with side information of Ahlswede-Korner and Wyner \cite{AhlswedeKorner1975}, \cite{WynerSC1975} where arbitrary distortion is allowed for one of the sources and the other source should be reconstructed losslessly. Finally, Beger and Yeung \cite{BergerYeung1989} considered a setting where one of the sources is to be perfectly reconstructed and the other source should be reconstructed with a distortion constraint (their setting subsumes all previous examples).
Recent results for specific sources or distortion measures include the achievable rate distortion region for the quadratic Gaussian multiterminal source coding problem, given by Wagner, Tavildar, and Viswanath  \cite{WagnerTavildarVisw2008} and the characterization of the region under logarithmic loss given by Courtade and Weissman \cite{CourtadeWeissman2011}. 

We consider a zero-delay version of the multiterminal source coding problem. We show that, unlike the arbitrary delay setting, when the zero delay constraint is imposed, this problem can be readily solved and the rate distortion region can be characterized. 

\subsection{Formal Definition and Main Result}
\begin{figure}[htp]
\centering
\includegraphics[width=0.9\textwidth]{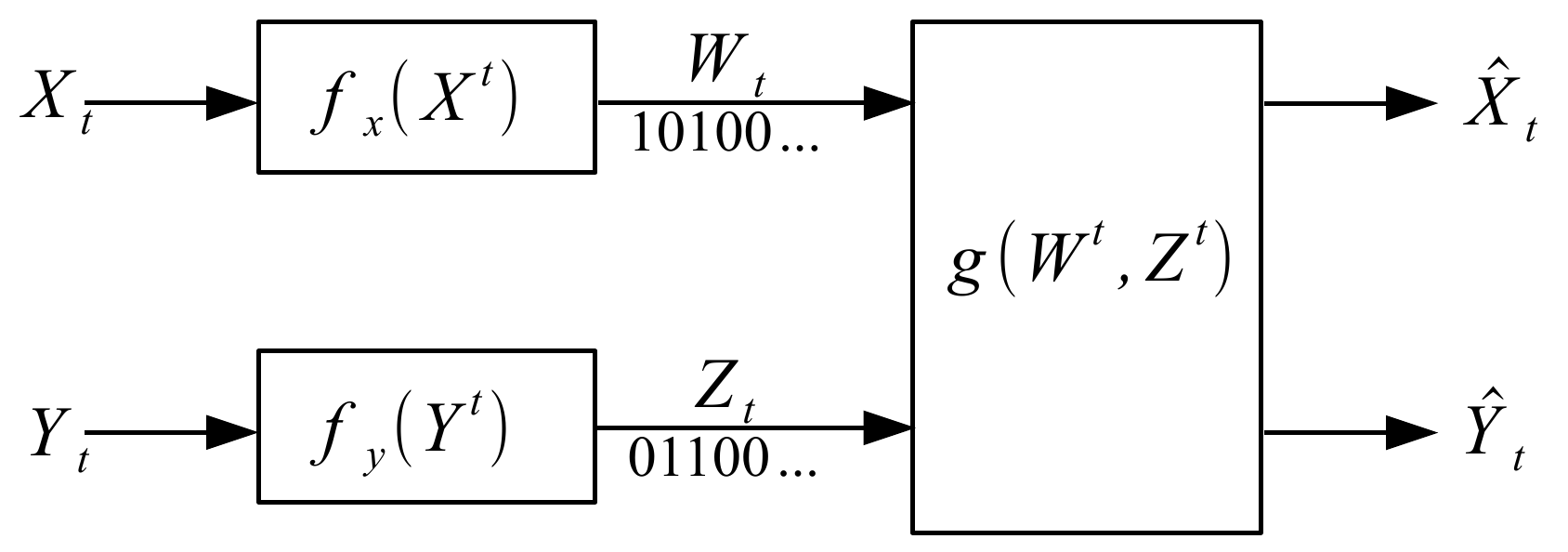}
\caption{Two users system model} \label{Fig:MultiTerm}
\end{figure}

We start by formally stating the problem. Let $(X_t,Y_t)$ be emitted by a memoryless source. Each of the symbols are encoded by $f_t^x,f_t^y$ respectively to produce $W_t=f_{x,t}(X^t), Z_t=f_{y,t}(Y^t)$. A decoder, $g_t: \calW^t\times\calZ^t\to\hat{\calX}\times\hat{\calY}$, uses $(W^t,Z^t)$ to reproduce the pair $(\hX_t,\hY_t)$. Let $(g_{x,t},g_{y,t})$ denote the reproduction functions of $X,Y$ respectively. Given two distortion measures, $d_x:\calX\times\hat{\calX}\to\mathbbm{R}^+,d_y:\calY\times\hat{\calY}\to\mathbbm{R}^+$, it is required that $\frn\sum_{t=1}^n \bE d_x(X_t,\hX_t)\leq D_x$ and $\frn\sum_{t=1}^n \bE d_y(Y_t,\hY_t)\leq D_y$ simultaneously. Each encoder encodes its message with an instantaneous code. As noted in the Introduction, there is more than one way to define the zero-delay decoding in a multiterminal setting. We allow either $Z_t$ or $W_t$ to be decoded first by the decoder and therefore the code of the later symbol (and generally all past decoded symbols) can serve as SI. Let $l_t^x, l_t^y$ denote the number of bits transmitted until the time that $X_t$ and $Y_t$ are encoded (inclusive), respectively.  The system model is depicted in Fig.\ref{Fig:MultiTerm}.

Let $\Xi$ denote the subset of $\mathbbm{R}^4$ consisting of all quadruples  $(R_x,R_y,D_x,D_y)$ such that for any given $\epsilon>0$ there exists a sequence functions $\cbr{f_{x,t}},\cbr{f_{y,t}},\cbr{g_t}$ and an integer $n$ such that
\begin{enumerate}
 	\item[i.] $\frn \bE l_n^x\leq R_x+\epsilon$ 
	\item[ii.] $\frn \bE l_n^y\leq R_y+\epsilon$
	\item[iii.] $\frn \sum_{t=1}^n \bE d_x(X_t,\hX_t)\leq D_x+\epsilon$
	\item[iv.] $\frn \sum_{t=1}^n \bE d_y(Y_t,\hY_t)\leq D_y+\epsilon$
\end{enumerate}
 
Our goal is to characterize the above operationally defined region by single-letter information-theoretic functionals. Towards this goal, 
let us define $\Delta^*_{YX}$ to be the set containing quadruples $(R_x,R_y,D_x,D_y)$ such that there exists random variables $(U,V,Q)$ that belong to finite alphabets $\calU,\calV,\calQ$, $|\calQ|\leq 5$, $|\calU|\leq |\calX|\times|\calQ|$,$|\calV|\leq |\calY|\times|\calQ|$ which are jointly distributed with $(X,Y)$ and satisfy
\begin{enumerate}
 	\item $U =  f_x(X,Q), V=f_y(Y,Q)$, for some functions $f_x:\calX\times\calQ\to\calU,f_y:\calY\times\calQ\to\calV$, $Q$ is independent of $(X,Y)$.
	\item $\exists g_x,g_y \text{ such that } g_x:\calU\times\calV\times\calQ\to\hat{\calX}, g_y:\calU\times\calV\times\calQ\to\hat{\calY}$ satisfying $\bE d_x(X,g_x(U,V,Q))\leq D_x$ and $\bE d_y(Y, g_y(U,V,Q))\leq D_y$. 
 	\item $R_x \geq L_{V}(U|Q)$\label{item_Rx}
	\item $R_y \geq L(V|Q)$\label{item_RY}
\end{enumerate}
The subscript in $\Delta^*_{YX}$ stands for the order of transmission, where the message of the $Y$-encoder, $V$, is decoded first and serves as SI to the message from the $X$-encoder. We similarly define $\Delta^*_{XY}$ in the same way but with reversed transmission order, i.e., items \ref{item_Rx},\ref{item_Ry} is defined as $R_x \geq L(U|Q)$ and $R_y\geq L_U(V|Q)$. Finally let $\Delta^* = \Delta^*_{XY}\cup \Delta^*_{YX}$  and define $\overline{\Delta^*}$ to be the closure of $\Delta^*$. 

The main results of this section is the following theorem:
\begin{theorem}\label{Thm:MTSC}
	$\Xi = \overline{\Delta^*}$. 
\end{theorem}

Theorem \ref{Thm:MTSC} basically states that using scalar encoders and decoders is optimal. The role of the encoder that transmits first is to send a message that will not only allow the joint reconstruction, but also to serve as SI for the decoding of the other message, thus reducing the average length which is needed for the other message. In the proof of the converse, we will show that there is no performance gain if each of the encoders receive the observations of the other encoder with delay (without delay, this becomes a single user problem for the source $(X,Y)$ with two distortion measures).  
There are several obvious extensions to this problem. For example, We can consider breaking each user's message into small pieces and consider all protocols that order the transmission of the pieces between the users, where all previously transmitted pieces serve as SI. Then, we can optimize over all possible such protocols (since the alphabets of $X$ and $Y$ are finite, there is a finite number of such protocols). Additionally, SI can be available to the decoder. Also, the encoder that transmits first might share some information with the second encoder at a given rate. The extension to such scenarios follows the same lines of proof as those of Theorems \ref{Thm:RTRD}-\ref{Thm:MTSC} and is therefore omitted. Another variant of this problem is a two-way problem where the $X$-encoder and $Y$-encoder communicate, where each user reproduces the other source to within a given distortion. It can be shown that in such a problem the past is, again, irrelevant and scalar encoders are optimal. In such a scenario, the interaction can be used both to assist the reconstruction and to reduce the average length which is needed to convey each message, as shown by Orlitsky in \cite{Orlitsky1992}.

We prove the converse and direct parts in the next two subsections, respectively.


\subsection{Converse:}
We prove the converse assuming $Z_t$ is sent first. The proof when $W_t$ is sent first is the same.
Once again, we use a ``genie aided'' scheme, where each encoder has access to $(X^{t-1}, Y^{t-1})$ when encoding $X_t$ or $Y_t$. We will show that even when each encoder knows the past symbols of the other source, still scalar encoders are optimal. With this scheme the only element that is unknown to each of the encoders is the current message of the other encoder and the decoder's data is shared by both encoders. 

For any sequence of encoders $\cbr{f_{x,t}(X^t,Y^{t-1})}$, $\cbr{f_{y,t}(X^{t-1}, Y^t)}$ and decoders $\cbr{g_{x,t}(W^t,Z^t})$, $\cbr{g_{y,t}(W^t,Z^t})$ achieving $(R_x,R_y,D_x,D_y)\in \Xi$ the following holds:
\begin{align}
 	nR_y &\geq \sum_{t=1}^n L(Z_t|X^{t-1},Y^{t-1})\nl
	&= \sum_{t=1}^n L(f_{y,t}(X^{t-1},Y^{t-1}, Y_t)|X^{t-1},Y^{t-1})\nl
	&= \sum_{t=1}^n L(f_{y,t}(\tilde{Q}_t, Y_t)|\tilde{Q}_t) \label{eq:MUR_y}
\end{align}
where we set $\tilde{Q}_t = (X^{t-1},Y^{t-1})$, which is independent of $(X_t,Y_t)$. We continue by letting $M$ be a random variable distributed uniformly over $\cbr{1,2,\ldots,n}$, independent of all other variables. Continuing \eqref{eq:MUR_y} we have
\begin{align}
 	R_y &\geq \sum_{t=1}^n L(f_{y,t}(\tilde{Q}_t, Y_t)|\tilde{Q}_t) \nl
	&= \frn\sum_{t=1}^n  L(f_{y,t}(\tilde{Q}_t, Y_t)|\tilde{Q}_t, M=t)\nl
	&=  L(\tilde{f}_{y}(M,\tilde{Q}_M, Y_M)|\tilde{Q}_M, M)\nl
	&=  L(\tilde{f}_{y}(Q, Y)|Q)\nl
	&=  L(V|Q)
\end{align}
where $Q\eqd(\tilde{Q}_M, M)$,  and $V \eqd \tilde{f}_{y}(Q, Y)$ and we used the fact that $Y$'s distribution does not depend on the time.

Using the same definitions of $M, \tilde{Q}_t, Q,V$ and following the same steps, we have for the rate of the $X$-encoder: 
\begin{align}
 	R_x &\geq\frn \sum_{t=1}^n L_{Z_{t}}(W_t|X^{t-1},Y^{t-1})\nl
	&= \frn\sum_{t=1}^n  L_{f_{y,t}(X^{t-1},Y^{t-1},Y_t)}(f_{x,t}(X^{t-1},Y^{t-1},X_t)|X^{t-1},Y^{t-1})\nl
	&= \frn\sum_{t=1}^n  L_{f_{y,t}(\tilde{Q}_t,Y_t)}(f_{x,t}(\tilde{Q}_t,X_t)|\tilde{Q}_t)\nl
	&= \frn\sum_{t=1}^n  L_{f_{y,t}(\tilde{Q}_t,Y_t)}(f_{x,t}(\tilde{Q}_t,X_t)|\tilde{Q}_t,M=t)\nl
	&=  L_{\tilde{f}_y(M, \tilde{Q}_M,Y_M)}(\tilde{f}_x(M, \tilde{Q}_M,X_M)|\tilde{Q},M)\nl
	&=  L_{\tilde{f}_y(Q,Y)}(\tilde{f}_x(Q,X)|Q)\nl
	&= L_{V}(U|Q)
\end{align}
where $U\eqd \tilde{f}(Q,X)$.

For the distortion we have that:
\begin{align}
	&D_x \geq  \frn\sum_{t=1}^n d_x(X_t, \hX_t) \nl
	&= \frn\sum_{t=1}^n \bE d_x(X_t, g_{x,t}(W^t,Z^t))\nl
	&\geq \frn\sum_{t=1}^n  \bE\sbr{d_x(X_t, \tilde{g}_{x,t}(X^{t-1},Y^{t-1}, f_{x,t}(X^{t-1},Y^{t-1},X_t), f_{y,t}(X^{t-1},Y^{t-1},Y_t)))}\label{eq:MU_DConv1}\\
	&= \frn\sum_{t=1}^n \bE\sbr{d_x(X_t, \tilde{g}_{x,t}(\tilde{Q}_t, f_{x,t}(\tilde{Q}_t,X_t), f_{y,t}(\tilde{Q}_t,Y_t)))}\nl
	&= \frn\sum_{t=1}^n \bE\sbr{d_x(X_t, \tilde{g}_{x,t}(\tilde{Q}_t, f_{x,t}(\tilde{Q}_t,X_t), f_{y,t}(\tilde{Q}_t,Y_t)))|M=t}\nl
	&= \bE\sbr{d_x(X_M, \tilde{g}_x(M,\tilde{Q}_M, \tilde{f}_x(M,\tilde{Q}_M,X_M), \tilde{f}_y(M,\tilde{Q}_M,Y_M)))}\nl
	&= \bE\sbr{d_x(X, \tilde{g}_x(Q, U,V))}
\end{align}
where in \eqref{eq:MU_DConv1} we used the fact that giving the decoder access to $(X^{t-1}, Y^{t-1})$ can only decrease the average distortion. In exactly the same manner, by replacing the roles of $X$ and $Y$, we get
\begin{align}
	D_y \geq \bE\sbr{d_y(Y, \tilde{g}_y(Q, U,V))}.
\end{align}

The bound on the cardinality of $Q$ is a consequence of Caratheodory's theorem \cite{Csiszar}:  As can be seen in \eqref{eq:MTq1}-\eqref{eq:MTq4}, the point 
$$(L_V(U|Q),L(V|Q),\bE d_x(X,g_x(U,V,Q)), \bE d_y(X,g_y(U,V,Q)))$$
lies in the convex hull of the set $\cbr{L_V(U|Q=q), L(V|Q=q), D_x(q), D_y(q)}$ 
which is a subset of $\mathbbm{R}^4$. 
Therefore, by Caratheodory's theorem, this region will not change if we restrict the alphabet size of $Q$ to 5.

\subsection{Direct:} 
In order to show that $\overline{\Delta^*} \subseteq \Xi$ we timeshare scalar quantizers. For every quadruple $(R_x,R_y,D_x,D_y)\in\overline{\Delta^*} $ We have
\begin{align}
 	L_V(U|Q) &= \sum_{q=1}^{|\calQ|} P(Q=q)L_V(U|Q=q)  \nl
	&\eqd  \sum_{q=1}^{|\calQ|} P(Q=q)R_x(q),\label{eq:MTq1}\\
	L(V|Q) &= \sum_{q=1}^{|\calQ|} P(Q=q)L(V|Q=q) \nl
	&\eqd   \sum_{q=1}^{|\calQ|} P(Q=q)R_y(q), \label{eq:MTq2}\\
	\bE d_x(X, g(U,V,Q)) &= \sum_{q=1}^{|\calQ|}P(Q=q)\bE\sbr{ d_x(X, g_x(U,V,Q))|Q=q}\nl
	&\eqd\sum_{q=1}^{|\calQ|}P(Q=q)D_x(q), \label{eq:MTq3}\\
	\bE d_x(X, g(U,V,Q)) &= \sum_{q=1}^{|\calQ|}P(Q=q)\bE\sbr{ d_y(Y, g_y(U,V,Q))|Q=q}\nl
	&\eqd\sum_{q=1}^{|\calQ|}P(Q=q)D_y(q).  \label{eq:MTq4}
\end{align}
Now, for every $q\in \calQ$, we will use the scalar encoders $f_x(X,q), f_y(Y,q)$ which define $U$ and $V$ (using rates $R_x(q), R_y(q)$ respectfully) and the scalar decoders $g_x(U,V,q)$ and $g_y(U,V,q)$ to reconstruct $\hX,\hY$ with average distortions $D_x(q)$ and $D_y(q)$ respectfully. Now for $n$ large enough, we can find $n_1,n_2, \dots, n_{|\calQ|}$ such that $\sum_{q=1}^{|\calQ|} n_q=n$ and $n_q/n$ is arbitrarily close to $P(Q=q)$. For each $q\in\calQ$, we use the encoder and decoders pertaining to $Q=q$ for $n_q$ coding stages. With this timesharing we can approach $(R_x+\epsilon,R_y+\epsilon,D_x+\epsilon,D_y+\epsilon)$ for any $\epsilon > 0$ if $n$ is large enough. 

\section{Conclusions}\label{Sec:Conclusion}
In this work we derived the fundamental limits of zero-delay lossy source coding with SI. It was shown that for the single-user setting, results in the spirit of \cite{NeuhoffGilbert1982} continue to hold in the sense that time-sharing scalar quantizers, followed by SI aware encoders is optimal. These codes have the feature that increasing the encoder's alphabet size can reduce the average length and not the more intuitive opposite, which is true when SI is not available and Huffman codes are used. Furthermore, we showed that in the multiterminal setting, the zero-delay variant of the multiterminal source coding problem can be readily solved, unlike the arbitrary-delay problem, which is open for  three decades. We also discussed possible extensions which can be proved following the same methods we used in this paper. Although we discussed only finite alphabets, we believe that extending the results of this paper to continuous alphabets in the multiterminal setting or continuous alphabets with discrete SI (for example if the decoder has another quantized version of the source as SI) in the single-user setting should be straightforward. 

It was shown that without SI, there are cases where there is no loss in forcing causal/sequential decoding, compared to standard rate-distortion decoding which waits for the whole block to arrive before producing the reconstruction. In other all other cases, we showed that the redundancy, compared to optimal rate-distortion performance is less than one bit when variable-rate coding is allowed. We used a simple scheme that relied on the classical random rate-distortion codebook construction, and only changed the way the encoder and decoder operate. However, trying to do the same when SI is available results in a substantial performance degradation. It is an interesting question whether there is an unavoidable loss when restricting to sequential decoding in these cases or whether our scheme, which works without SI, is too naive when SI is available. 
\pagebreak
\appendix
\noindent\huge{Appendix}

\normalsize
\setcounter{equation}{0}
\numberwithin{equation}{section}
\section{Details of the Sequential Scheme}\label{App:Scheme}
In this appendix, we show the analysis of the sequential scheme which was briefly described in Section \ref{Sec:MainRes}. This scheme uses a classical rate-distortion random code \cite{cover}. The encoder finds the first codeword, $\hX^n$, in the codebook which is distortion-typical (as defined in \cite{cover}) and starts to transmit the \textit{reproduction symbols}, $\hX_t$, sequentially, using $\log|\hat{\calX}|$ bits each transmission (in contrast to the classical encoder which will send the index of the codeword in a block). The decoder outputs $\hX_t$ as it receives it. 
The idea here, is that after receiving $n\cdot\alpha$ symbols, the decoder can detect the specific codeword in the codebook since, with high probability, no other codeword will have the same prefix and the rest of the reproduction symbols ($\hX_{n\alpha+2}^n$) can be reproduced without further transmissions from the encoder. 

%
%
%
We start by analyzing the probability that given a sequence, if we now randomly draw a codebook, we will draw another sequence with the same prefix. Formally, Let $\hX$ be drawn with an i.i.d source $P(\hx)$. We first draw a single sequence of length $n$ according to $P(\hat{\bx})=\prod_{t=1}^nP(\hx_t)$, then we independently draw $2^{nR}$ sequences using the same source. We ask what is the probability that none of the $2^{nR}$ sequences start with the same $n\alpha$ symbols prefix of the first sequence? We average this over the probability to draw a specific sequence in the first drawing. If we found an $\alpha$ for which this probability vanishes as $n$ grows, this will give us a bound on the number of symbols we need to send in our sequential scheme in order to identify the correct codeword. 
Let this probability be denoted by $P_c$. We will treat only the first $n\alpha$ symbols since we do not care about the contents of the rest. We use the method of types. $P_{\bx}$ will denote the empirical distribution of $\bx$, i.e., $P_{\bx}(x)=N_{\bx}(x)/{n\alpha}$ where $N_{\bx}(x)$ counts the number of occurrences of the symbol $x$ in $\bx$. $H(P_{\bx})$ will denote the empirical entropy of the empirical distribution $P_{\bx}(\cdot)$. We use standard results of the method of types which can be found in \cite{Csiszar}. 

After we have the first sequence, the probability that it will not be the outcome of the next drawing is $1-2^{-n\alpha\rbr{D(P_{\bx}||P)+H(P_{\bx})}}$. Therefore,
\begin{align}
 	P_c &\exe \sum_{\bx}P(\bx)\rbr{1-2^{-n\alpha\rbr{D(P_{\bx}||P)+H(P_{\bx})}}}^{2^{nR}}\nl
	&\exe \sum_{P_{\bx}} 2^{-nD(P_{\bx}||P)}\rbr{1-2^{-n\alpha\rbr{D(P_{\bx}||P)+H(P_{\bx})}}}^{2^{nR}}\nl
	&\geq \rbr{1-2^{-n\alpha H(P)}}^{2^{nR}}
\end{align}
where the second equality is true since the the expression in the first line depends only on the type-class of $\bx$ and the last inequality is true since from the sum over all type-classes, we took only $P_{\bx}=P$ (we assume that $n$ is large enough so we can get as close as we want to $P$, even if it is not rational). The last expression converges to unity double exponentially fast if $\alpha H(P) > R$ or equivalently, $\alpha > \frac R {H(P)}$. 

We now use this result to analyse a coding system with sequential decoders. We draw a codebook consisting of $2^{n(R(D)+\epsilon)}$ codewords according to the prior on $\hX$ that is calculated from the channel that achieves the (standard) rate distortion function. Given a source sequence, $X^n$,  to be encoded, the encoder looks for a sequence in the codebook, $\hX^n$, that is distortion typical (see \cite{cover}) with the source sequence and starts sending the reproduction symbols $\hX_t, ~~t=1,2,\ldots,n\alpha$, sequentially, $\log|\hat{\calX|}$ bits at a time. The decoder outputs the reproduction symbols sequentially as it receives their description. With the above result, if $\alpha = \frac {R(D)} {H(\hX)}+2\epsilon$, there is no other codeword in the codebook with the same prefix with probability that converge to unity double exponentially fast. Therefore, after receiving $n\rbr{\frac{R(D)} {H(P)}+2\epsilon}\log|\hat{\calX|}$ bits, with very high probability, the decoder knows the whole sequence $\hat{X}^n$ and can output the rest. For the whole $n$-block, we send no more than $\frn  \sbr{n\rbr{\frac {R(D)} {H(\hX)}+2\epsilon}}\log|\hat{\calX|}$ bits per source symbol. This means that whenever the rate-distortion achieving prior is uniform, we achieve the optimal rate-distortion performance with this simple sequential scheme (since $H(\hX)=\log|\hat{\calX}|$). Otherwise, we are away from optimality by a factor of $\frac {\log|\hat{\calX}|} {H(\hX)}$. This factor can be reduced if we allow variable rate coding, i.e., relax the constant number of bits per transmission constraint to a constraint on the average number of bits per transmission. If variable rate coding of $\hX_t$ is allowed, we can upper bound the average number of bits we send each transmission by $H(\hX)+1$. In this case the the number of bits per source symbol we send is upper bounded by $R(D)+\frac {R(D)} {H(\hX)}\leq R(D)+1$, meaning we are less than one bit away from optimal performance with this simple sequential scheme for any source and distortion measure. 
Note that, essentially, we used the same random code used in the classical analysis of the achievability of  the rate-distortion results. The error event is the union of the events that the random code will not cover a specific source sequence with distortion $D$ and the event that the codeword that describes this source sequence with distortion $D$ has another codeword with the same prefix. The complete analysis that shows that there exists a codebook for which our scheme will work is the same as the one found in \cite[Chapter 10.5]{cover} and is therefore omitted. 
\section{Proof of Theorem \ref{Thm:ScalarDec}}\label{App:Scalar_Enc_Proof}
In this appendix, we prove the first part of Theorem \ref{Thm:ScalarDec}, namely, we show that $\calR_{s-e}(D) = \underline{R}_{ZD}(D)  
$. Once again, the direct part is the same as the direct part of Theorem \ref{Thm:RT_Y_LH}, we need to prove only the converse.

When the encoder is scalar (memoryless), it cannot use the past transmitted messages as SI. Therefore, for each message, at least $L_{Y_t}(W_t)$ are sent. We have 
\begin{align}
 	nR & \geq\sum_{t=1}^n \int L_{Y_t}(f_t(X_t))\nl
	& \geq\sum_{t=1}^n \int L_{Y_t}(f_t(X_t)|W^{t-1},W_{t+1}^n, Y^{t-1}, Y^{n}_{t+1})\nl
	&\geq\sum_{t=1}^n \int L_{Y_t}(f_t(X_t)|w^{t-1},w_{t+1}^n, y^{t-1}, y_{t+1}^n)d\mu(w^{t-1},w_{t+1}^n, y^{t-1}, y_{t+1}^n)\nl
	&\geq\sum_{t=1}^n \int L_{Y_t}(f_t(X_t))d\mu(w^{t-1},w_{t+1}^n, y^{t-1}, y_{t+1}^n)
\end{align}
Now, the rate in the inner expression cannot be smaller than the rate of the optimal scalar system (with all conditioned elements serving as index of functions), that achieves the same distortion as the given system (with decoder functions $h_t(w^{t-1},W_t,y^{t-1},Y_t,y_{t+1}^n)$
\begin{align}
	nR&\geq\sum_{t=1}^n \int L_{Y_t}(f_t(X_t))d\mu(w^{t-1},w_{t+1}^n, y^{t-1}, y_{t+1}^n)\nl
	&\geq \sum_{t=1}^n \int \underline{R}_{ZD}\left(\bE \left[d(X_t, g_t(f_t(X_t),w^{t-1}, w_{t+1}^n, y^{t-1},Y_t,y_{t+1}^n))\right]\right)\times\nl
	&~~~~~~d\mu(w^{t-1},w_{t+1}^n, y^{t-1}, y_{t+1}^n)\nl
\end{align}
From here, using the concavity and monotonicity of $\underline{R}_{ZD}(\cdot)$ and following the steps of the proof of Theorem \ref{Thm:RT_Y_LH} the proof will be concluded.

\bibliographystyle{IEEEtran}
\bibliography{../PhDBib-1}

\end{document}